\documentclass[a4paper]{article}
\usepackage[centertags,intlimits]{amsmath}
\usepackage{times}
\newcommand{\bea}{\begin{eqnarray}}
\newcommand{\bee}{\end{eqnarray}}
\newcommand{\bean}{\begin{eqnarray*}}
\newcommand{\been}{\end{eqnarray*}}
\begin{document}
\begin{flushright}
IISc-CTS-12/00 \\[2mm]
FZJ-IKP(Th)-2000-18
\end{flushright}

\vskip 1cm

\begin{center} {\large{\bf 
Comparison of \boldmath{$\pi K$} Scattering in \boldmath{$SU(3)$} Chiral
Perturbation Theory\\
and Dispersion Relations\footnote{Work supported
  in part by DFG under contract no. ME 864-15/2.}}}

\vskip 2cm
{\bf B. Ananthanarayan} \\
Centre for Theoretical Studies \\
Indian Institute of Science \\
Bangalore 560 012, India

\vskip 1cm

{\bf P. B\"uttiker} \\
Institut f\"ur Kernphysik \\
Forschungszentrum J\"ulich \\
D - 52425 J\"ulich, Germany
\end{center}

\vskip 1cm

\begin{abstract}
We establish the framework for the comparison
of $\pi K$ scattering amplitudes from $SU(3)$ chiral perturbation
theory with suitable dispersive representations which
result from the combination of certain fixed-t 
dispersion relations with dispersion relations on hyperbolic curves.  
This allows for predictions for some combinations of low energy
constants appearing in higher order calculations of chiral
perturbation theory. Using a simple parametrization for the lowest
partial waves, first estimates for some combinations are presented.
\end{abstract}

\bigskip


\medskip

\noindent P. A. C. S. Nos.: 12.39.Fe, 11.55.Fv, 13.75.Lb

\newpage

\section{Introduction}
The pseudoscalar octet of pions, kaons and the $\eta$ may be
viewed as the Goldstone bosons of the spontaneously broken
approximate symmetry of the QCD Lagrangian whose interactions
may be described by $SU(3)$ chiral perturbation theory~\cite{GLNPB}.
The $\pi K$
scattering amplitudes have been computed in this framework sometime ago, 
see, ref.~\cite{BKM,dobado}. For an extensive review of phenomenological
information prior to these developments, including dispersion 
relation analysis, we refer to ref.~\cite{Lang}.  
Our aim here is to set up the appropriate framework within which 
the chiral amplitudes  can be compared with dispersive 
representations of the amplitudes,
of the type established in axiomatic field theory.

It is instructive to first discuss $\pi\pi$ 
scattering which has been studied in considerable detail.
The $\pi \pi$ scattering amplitude has been computed to 
one-loop accuracy~\cite{GLANN}, and to 
two loops in $SU(2)$ chiral perturbation theory~\cite{KMSF1,BCEGS}, 
and at one-loop accuracy in $SU(3)$ chiral perturbation theory~\cite{BKM}.  
In $SU(2)$ chiral perturbation theory  up to two-loop accuracy, the
amplitude is described by three functions of a single (Mandelstam)
variable, whose absorptive parts are given in terms of those of the
three lowest partial waves~\cite{SSF}.   One replaces them by an
dispersive representation which yields an amplitude with
an effective low energy polynomial and a dispersive tail~\cite{KMSF1,AB1}.
A dispersion relation representation with two subtractions
is an ideal starting point for rewriting them in a form
whereby a comparison with the chiral representation can be made,
when the $S$-- and $P$--wave absorptive parts alone are retained. The
absorptive parts of the higher waves contribute to the 
polynomial pieces only.  We note that the analysis in the
past has been performed only for elastic scattering in $SU(2)$ chiral perturbation theory;
we extend it in a straightforward manner to $SU(3)$ chiral perturbation theory where
$K\overline{K}$ and $\eta \eta$  are present in intermediate states.  
The remarkable synthesis of dispersion relation phenomenology
(see \cite{acgl} and references therein) and
chiral perturbation theory has recently led to a highly accurate
prediction for the iso-scalar $S$--wave scattering length\cite{clg}.

Recently, there has been an revival in the interest in $\pi K$ scattering.
There are indications for a flavour dependence of the size of the quark
condensate \cite{Stern}. As $\pi K$ scattering is the most simple
$SU(3)$--process involving kaons, it is the suitable place to test
this dependence.  Furthermore, there are plans to measure the
lifetime of $\pi K$--atoms to an accuracy of $20\%$ in the  DIRAC
experiment at CERN \cite{Dirac}. This
would allow for a very precise determination of the scattering length
$a^{-}_{0}$ within an error of $~ 10\%$.   The role of the latter must not
be understated: one may show that in $SU(3)$ chiral perturbation
theory to one loop the scattering length $a^-_0$ depends only on the
constant $L^r_5$ and the pion and the kaon decay
constant. Furthermore, $L^r_5$ is itself determined, at this order, by
the ratio of these two decay constants, see
\cite{GLNPB}, leaving $a^-_0$ free of any low energy constants.
This observation is being published here for the first time, although it
was already known \cite{ulf_remark} to the authors of
\cite{BKM}.
Note that a similar statement holds for the
tree--level prediction of $a^-_{0}$ in generalized chiral perturbation
theory, where the {\it a priori} unknown low energy constants appear even at
tree--level \cite{Knecht:1993eq}.
Therefore, a comparison of experimental values of
$a^{-}_{0}$ with its theoretical predictions from chiral perturbation
theory is a stringent test of the framework of chiral symmetry breaking.
Apart from that, an independent high precision estimate for this
scattering length calls for a fresh partial wave analysis since this
threshold parameter can play the role of a subtraction constant in $\pi
K$ dispersion relations.

The structure of the $\pi K$ amplitudes is best revealed when  
we consider a system of amplitudes 
defined by $T^+(s,t,u)$ and $T^-(s,t,u)$, which are even and odd 
under the interchange of $s$ and $u$, respectively.  
We demonstrate that each of these chiral amplitudes
may be written down in terms of three functions of single variables whose
absorptive parts are related to those of the $S$-- and $P$--waves in the
$s$-- and $t$--channels.  

The dispersion relations we employ 
for $T^{+}$ are the ones as given in ref.~\cite{JN1} 
where fixed-t dispersion relations are combined with dispersion
relations on hyperbolic curves. 
We introduce a new dispersion relation here for $T^-$.  
Retaining only the absorptive parts of the $S$-- and $P$--waves we demonstrate
the equivalence of the structure of the representation to that of the
chiral result.   This allows us to match the representations in the low energy
domain, after adding the contributions of the higher waves which
are only polynomials to this order. Since $\pi K$ scattering at
low energies is dominated by the $S$-- and $P$--waves, a detailed
knowledge of these waves is important.  Stringent constraints
resulting from axiomatic analyticity and crossing are best expressed
in terms of integral (Steiner-Roy) equations
\cite{Lang,Steiner,Roy,Karabarbounis}.  These equations are the ideal starting
point for a future analysis of $\pi K$ scattering   information. Due
to the importance of these equation in such an analysis, the
Steiner-Roy equations in the $S$- and $P$-wave approximation for the
$S$- and $P$-waves are given here explicitly. 
(It may be noted that one can proceed to analyze $\pi N$ scattering
\cite{AB3} in an analogous manner.)
 
While an accurate phase shift parametrization, independent from input
from chiral perturbation theory is awaited, we use a simple $K$-matrix
approach to saturate  the dispersion relation which then
provide the first estimates for certain combinations of the low energy
constants of $SU(3)$ chiral perturbation theory from $\pi K$ scattering. 

The plan of the paper is as follows:  in Sec.~2 we establish our
notation and conventions, in Sec.~3 we carry out the decomposition
of the one-loop $\pi K$ amplitudes into functions of single variables
and then discuss the method of replacing them with a dispersive
representation. In Sec.~4 the dispersion relations are considered and
rewritten in the $S$-- and $P$--wave absorptive part approximation,
the contributions of the higher waves are discussed, and the comparison
with the chiral amplitude is outlined. Furthermore, we explicitly give
the Roy equations for the lowest $\pi K$ partial waves. In Sec.~5 we
saturate the dispersion relations with phenomenological absorptive parts and 
discuss the consequences.  In Sec.~6 we provide a discussion and a
summary of the results. Appendix A  briefly
summarizes the results obtained when applying the matching mentioned
above to $SU(3)$ $\pi\pi$ scattering, in appendix B functions of
single variables of interest are listed, and in appendix C the kernels
of the Steiner-Roy integral equations for the $S$-- and $P$--waves are given.
\section{Notation and Conventions}
We consider the process
\bean
\pi^{I_1}(p_1) + K^{J_1}(q_1) \to \pi^{I_2}(p_2) + K^{J_2}(q_2),
\been
with the four-momenta $p_i, q_i$ and the isospin $I_i$ and $J_i$ of
the pions and the kaons, respectively. The Mandelstam variables are defined
as ($\Sigma\equiv M^2 + m^2$)
\begin{eqnarray*}
   s  =  (p_1 + q_1)^2,\,
   t  =  (q_1 - q_2)^2,\,
   u  =  (q_1 - p_2)^2,
\end{eqnarray*}
with
\begin{eqnarray*}
   s + t + u = 2 \Sigma,
\end{eqnarray*}
where $M$ and $m$ are the pion and the kaon mass, respectively.
In the $s$-channel the center of mass scattering angle
$\Theta_s$ and momentum $q_s$ are given by ($\Delta \equiv M^2 - m^2$)
\begin{eqnarray*}
    z_s  &\equiv&  \cos\Theta_s = 1 + \frac{t}{2 q_s^2} =
                                      \frac{t-u+\frac{\Delta^2}{s}}{4 q_s^2},\\
\end{eqnarray*}
\begin{eqnarray*}
   q_s^2 & = & \frac{(s-(m-M)^2)(s-(m+M)^2)}{4 s},
\end{eqnarray*}
and the partial wave decomposition is defined by
\bean
   T^{I_s}(s,t,u) & = & 16 \pi \sum (2 l + 1) f^{I_s}_l(s) P_l(z_s).
\been

The partial waves may then be parametrized by the phase shifts
$\delta^{I}_{l}$ and the inelasticities $\eta^{I}_{l}$,
\bean
   f^{I}_{l}(s) & = & \frac{\sqrt{s}}{2 q_s}\frac{1}{2
     i}\left\{\eta^{I}_{l}(s)e^{2 i \delta^{I}_{l}(s)}-1\right\},
\been
and have the threshold expansion
\bean
   {\rm Re}\, f^I_l(s) = \frac{\sqrt{s}}{2}q^{2 l}\left\{ a^I_l +
     b^I_l q^2 + O(q^4)\right\}.
\been
In the $t$-channel, the center of mass momenta of the pion and the
kaon are  $q_t$ and $p_t$, respectively, and the centre of
mass scattering angle $\Theta_t$ is given by
\begin{eqnarray}\label{eq:t_ch_kin}
   z_t & \equiv & \cos\Theta_t = \frac{s+p_t^2+ q_t^2}{2 q_t p_t} =
   \frac{s-u}{4 p_t q_t},\nonumber\\
   p_t  & = & \sqrt{\frac{t - 4 m^2}{4}},\,
   q_t   =  \sqrt{\frac{t - 4 M^2}{4}}. \nonumber
\end{eqnarray}
The partial waves are defined by
\bean
   T^{I_t}(s,t,u) & = & 16 \pi \sqrt{2}\sum (2 l + 1) f^{I_t}_l(t) P_l(z_t).
\been
Once one of the isospin amplitudes is known the other and
  combinations of these are fixed by crossing symmetry:
\bean
   T^{1/2}(s,t,u) & = & \frac{3}{2}T^{3/2}(u,t,s) - 
                        \frac{1}{2}T^{3/2}(s,t,u),\\
   T^{+}(s,t,u)  & \equiv & \frac{1}{3}T^{1/2}(s,t,u) +
                           \frac{2}{3}T^{3/2}(s,t,u), \, 
                 =       \frac{1}{\sqrt{6}} T^{I_t=0}(s,t,u),
                           \label{eq:Tplus}\\
   T^{-}(s,t,u)  & \equiv & \frac{1}{3}T^{1/2}(s,t,u) -
                           \frac{1}{3}T^{3/2}(s,t,u) \, 
                 =       \frac{1}{2}T^{I_t=1}(s,t,u)
\been
It may be seen from the above that $T^+(s,t,u)$ is even under the interchange
of $s$ and $u$, whereas $T^-(s,t,u)$ is odd.
\section{Decomposition of the chiral amplitudes}
In the framework of one-loop $SU(3)$ chiral perturbation theory, the explicit
expression for $T^{3/2}(s,t,u)$ has been presented in ref.~\cite{BKM}.
One then constructs the two amplitudes of interest $T^+(s,t,u)$
and $T^-(s,t,u)$.   It may be seen that these can now be decomposed into
terms involving functions of single variables only:
\begin{equation}
\begin{split}\label{eq:chiral_decomp}
   T^+(s,t,u) &=  Z^{+}_0(s) + Z^{+}_0(u)
               + (t-s+\frac{\Delta^2}{u}) Z^{+}_1(u) + 
                    (t-u+\frac{\Delta^2}{s})Z^{+}_1(s) + Z^{+}_t(t),\\
   T^-(s,t,u) &=  Z^{-}_0(s) - Z^{-}_0(u)
               +(t-s+\frac{\Delta^2}{u}) Z^{-}_1(u)
                    -(t-u+\frac{\Delta^2}{s}) Z^{-}_1(s) + (s-u) Z^{-}_t(t) .
\end{split}
\end{equation}
Written in this form, 
the imaginary parts of the $Z$'s are related to those of
the lowest partial waves in the following manner ($s \geq (m+M)^2, t
\geq 4 M^2$):
\begin{equation}
\begin{split}\label{imageqs}
   {\rm Im\,} Z^{\pm}_0(s) & =  16 \pi \,{\rm Im\,}f^\pm_0 (s), \\
   {\rm Im\,} Z^{\pm}_1(s) & =  \frac{12 \pi}{q_s^2}{\rm Im\,}f^\pm_1
   (s), \\ 
   {\rm Im\,} Z^{+}_t(t) & =  \frac{16 \pi}{\sqrt{3}}{\rm
     Im\,}f^{I_t=0}_0 (t),\\
   {\rm Im\,} Z^{-}_t(t) & =  6\sqrt{2}\pi\, {\rm
     Im\,}\frac{f^{I_t=1}_1(t)}{p_t q_t}. 
\end{split}
\end{equation}
In appendix B we present our choice of $Z^\pm_i, \,
i=0,1,t$.\footnote{\label{fn:ambig_decomp}The decomposition does not
  uniquely fix the algebraic parts of the functions, which is a
  consequence of not all the Mandelstam variables being independent.}
The imaginary parts of $Z^\pm_0$ and $Z^\pm_1$ receive
contributions from the $\pi K$ and $K\eta$ loops with the lower cut starting at
the $\pi K$ threshold $s=(M+m)^2$.  On the other
hand the imaginary parts of $Z^\pm_t$ receive contributions from the
$\pi\pi$ and $K\bar{K}$ loops and $Z^+_t$ alone from $\eta\eta$ loops, 
with the lowest cut starting at the $\pi\pi$ threshold $s=4 M^2$.   
The former when written out in terms of amplitudes of definite iso-spin
in the $s-$ channel are such that they
respect the elastic unitarity condition
\bean \nonumber
{\rm Im} f^I_l(s) = \frac{2 q_s}{\sqrt{s}} |f^I_l(s)|^2.
\been
The latter respect the principle of extended unitarity, {\it viz.}
\bean \nonumber
{\rm arg} f^{I_t=0}_0(t) = \delta^0_0 (t)(\pi\pi),\quad{\rm arg}f^{I_t=1}_1(t)=
\delta^1_1(t) (\pi\pi),\quad 4 M^2 \leq t \leq 4 m^2.
\been
Keeping this in mind, and using eq.~(\ref{imageqs}), 
it can be shown that the $Z^\pm_i, \, i=0,1,t$ verify the
following relations (written out to enable a comparison of the chiral
and dispersive amplitudes to this order in chiral perturbation
theory):
\begin{equation}
\begin{split}
\label{eq:dispW_X}
   Z^{\pm}_0(s) & =  
                 \frac{\alpha^{\pm}_0}{s} + \beta^{\pm}_0 + \gamma^{\pm}_0 s +
                 \delta^{\pm}_0 s^2 + 
                16 s^3
                \int_{(m+M)^2}^{\infty}\frac{ds'}{{s'}^3}\frac{{\rm
                Im\,}f^\pm_0(s')}{s'-s}, \\
   Z^{\pm}_1(s) & =  \beta^{\pm}_1 + \gamma^{\pm}_1 s + 
                12 s^2
                \int_{(m+M)^2}^{\infty}\frac{ds'}{{s'}^2}\frac{1}
                {q_{s'}^2}
                \frac{{\rm
                Im\,}f^\pm_1(s')}{s'-s},\\
   Z^{+}_t(t) & =  \beta^{+}_t + \gamma^{+}_t t + \delta^{+}_t t^2 + 
                \frac{16 t^3}{\sqrt{3}}
                \int_{4 M^2}^{\infty}\frac{dt'}{{t'}^3}\frac{{\rm
                Im\,}f^{I_t=0}_0(t')}{t'-t},\\
      Z^{-}_t(t) & =  \beta^{-}_t + \gamma^{-}_t t + 
                6 \sqrt{2}t^2 \int_{4 M^2}^{\infty} \frac{dt'}{{t'}^2}
                \frac{1}{t'-t}{\rm Im\,} 
                \frac{f^{I_t=1}_{1}(t')}{p_{t'} q_{t'}}. 
\end{split}
\end{equation}
The subtraction constants $\alpha^{\pm}_i, \beta^{\pm}_i,
\gamma^{\pm}_i,$ and $\delta^{\pm}_i$ depend on the low energy
constants $L^r_i$ and may be simply evaluated from the
explicit expressions we have provided for the $Z^{\pm}_i,\,
i=0,1,t$. Note that the appearance of the poles in $Z^{\pm}_0$ is due
to the unequal masses of the particles. However, they cancel the
kinematic poles appearing in the coefficients of $Z^{\pm}_1$ such that
in the chiral representation (\ref{eq:chiral_decomp}) these poles disappear.
With eqs.~(\ref{eq:dispW_X}) and (\ref{eq:chiral_decomp}) we may write
\begin{equation}\label{eq:chiral_Tplus}
\begin{split}
   T^{+}(s,t,u) &= 2\,\beta^{+}_0 + \beta^{+}_t - 2\,\left( m^4 +
     6\,m^2\,M^2 + M^4 \right) \,\gamma^{+}_1 
                + \left( s+u \right) \,\left(
                  \gamma^{+}_0-\beta^{+}_1 \right)+\left( s^2 + u^2
                \right) \,   \left(\gamma^{+}_1 + \delta^{+}_0
                \right)\\ 
               &  + t\,\left( 2\,{\beta^{+}_1} +6\,\left( m^2 +
                    M^2 \right) \,\gamma^{+}_1 + {\gamma^{+}_t}
                \right) + t^2\,\left( {\delta^{+}_t} -
                  2\,{\gamma^{+}_1}\right)\\ 
                \, + 16\!\!\!\!\!&\int_{(m+M)^2}^{\infty}\!\!\!\!\frac{d\,s'}{{s'}^3}
                   \left[\frac{s^3}{s'-s}+\frac{u^3}{s'-u}\right]{\rm
                     Im} f^{+}_0(s') 
                   + \frac{16}{\sqrt{3}}t^{3}\int_{4
                     M^2}^{\infty}\frac{d\,t'}{{t'}^3} 
                   \frac{{\rm Im\,}f^{I_t=0}_0(t')}{t'-t}\\ 
                  +12\,&s^2
                \left(t-u+\frac{\Delta^2}{s}\right)\int_{(m+M)^2}^{\infty} 
                \!\!\!     \frac{d\,s'}{{s'}^2}\frac{1}{q_{s'}^2}
                     \frac{{\rm Im\,}f^{+}_{1}(s')}{s'-s}
                  +12\,u^2
                \left(t-s+\frac{\Delta^2}{u}\right)\int_{(m+M)^2}^{\infty} 
                \!\!\!     \frac{d\,s'}{{s'}^2}\frac{1}{q_{s'}^2}
                     \frac{{\rm Im\,}f^{+}_{1}(s')}{s'-u},
\end{split}
\end{equation}
and
\begin{equation}\label{eq:chiral_Tminus}
\begin{split}
   T^{-}(s,t,u) &= \left( \beta^{-}_1 + \beta^{-}_t + \gamma^{-}_0 \right)
                   \,\left( s - u \right)  + \left(\gamma^{-}_1 
                   +\gamma^{-}_t\right)\,t\,\left( s-u \right) + \delta^{-}_0
                   \,\left( s^2 - u^2 \right)\\
                &\quad+16\!\!\!\int_{(m+M)^2}^{\infty}\!\!\!\frac{d\,s'}{{s'}^3}
                   \left[\frac{s^3}{s'-s}-\frac{u^3}{s'-u}\right]{\rm
                     Im\,}f^{-}_0(s') 
                   + 6\,\sqrt{2}\,t^2(s-u)\int_{4 M^2}^{\infty}
                  \frac{d\,t'}{{t'}^2 (t'-t)}{\rm Im\,} 
                  \frac{f^{I_t=1}_1(t')}{q_{t'} p_{t'}}\\ 
                &\quad+12 s^2\left(t-u+\frac{\Delta^2}{s}\right)
                   \!\!\!\int_{(m+M)^2}^{\infty}\!\!\!\!\!\!\!
                   \frac{d\,s'}{{s'}^2}
                   \frac{1}{q_{s'}^2}\frac{{\rm Im\,}f^{-}_1(s')}{s'-s}
                -12 u^2\left(t-s+\frac{\Delta^2}{u}\right)
                   \!\!\!\int_{(m+M)^2}^{\infty}
                   \!\!\!\!\!\!\!\frac{d\,s'}{{s'}^2}
                   \frac{1}{q_{s'}^2}\frac{{\rm Im\,}f^{-}_1(s')}{s'-u}.
\end{split}
\end{equation}
The polynomial part of $T^{\pm}$ then reads\footnote{The explicit
  expressions for $\alpha^\pm_i, \beta^\pm_i, \gamma^\pm_i,$ and
  $\delta^\pm_i$ may be obtained from the authors.}, after eliminating
the ambiguity associated with the mass shell condition $s+t+u=2 \Sigma$ (see
footnote \ref{fn:ambig_decomp}):
\begin{equation}
\begin{split}
  T_{P,\chi}^+(s,t,u) & = \left\{2 \beta_0^+ + \beta^+_t + 2\left[\Delta^2 \gamma_1^+ 
                      - \Sigma(\beta^+_1 - \gamma_0^+) + \Sigma^2
                        ( \delta_0^+ - \gamma_1^+)\right]\right\}  \\
               &   + \left\{3 \beta_1^+ - \gamma_0^+ + \gamma_t^+ -
                        2\Sigma(\delta_0^+ - 2 \gamma_1^+)\right\}
                     t + \frac{1}{2}(\delta_0^+ - 3 \gamma_1^+ + 2
                     \delta^+_t) t^2  \\ 
               &   + \frac{1}{2}(\delta_0^+ + \gamma_1^+)
               (s-u)^2,\label{chiralpoly}\\ 
  T_{P,\chi}^-(s,t,u) & =  (\beta_1^- + \beta_t^- +\gamma_0^- + 2 \Sigma
                      \delta_0^-)(s-u) + (\gamma_1^-+\gamma_t^- -
                      \delta_0^-)(s-u)t.   
\end{split}
\end{equation}
It might be noted that chiral perturbation
theory could provide an accurate description of the $\pi K$ scattering
amplitude in the low-energy domain, if we could compare the representation
given above with a suitable representation provided by dispersion
relations, upon exploiting analyticity and crossing properties of the
amplitudes.  Furthermore, it is the six lowest partial waves that
essentially determine the low-energy structure completely and also fix the
low energy constants when the chiral and dispersive representations are
compared, up to some unknown subtraction constants, a role that is played
by the scattering lengths. These partial waves, in principle, are related
through analyticity and crossing by integral equations that are generated
by the dispersion relations for the amplitudes $T^+$ and $T^-$.  In the next
section it is precisely those dispersion relations which provide this
framework which are first set up and analyzed and then used to generate 
the system of integral equations.  When the absorptive parts of all
$l \geq 2$ waves are neglected, the system of equations is a closed system
of equations for these waves and imposing unitarity on the partial waves
constrains them further.  Such a system could be used in the future for
an analysis of presently available and forthcoming data to pin down the
scattering lengths within relatively small uncertainties and to determine
the low energy constants through a program of sum rules.

Note that the $\pi K$ scattering amplitude at tree-level in generalized
chiral perturbation theory has
been discussed in ref.~\cite{Knecht:1993eq} and in the heavy-kaon
effective theory~\cite{Roessl}. 
Our methods can be extended to analyze these theories as well.
\section{Dispersion relations for \boldmath{$\pi K$} scattering}
In field theory the scattering amplitudes $T^+$ and $T^-$
verify fixed-$t$ dispersion relations, under conventional assumptions
regarding the high energy behaviour, the former with two subtractions
and latter with none. In practice,  we have found that
in order to meet the requirements of matching the chiral expansion
with the axiomatic representation, dispersion
relations with two subtractions for $T^-$ as well prove to be
convenient.  In ref.~\cite{JN1}, the unknown $t$- dependent
subtraction function was eliminated by considering dispersion
relations on a certain hyperbola, $s\cdot u = \Delta^2$, resulting in
a representation that we find most suitable for our purposes.  
The primary reason for this is that it is the choice of comparing the fixed-$t$
dispersion relations and the hyperbolic dispersion relations on
the hyperbola given above and at $t=0$ which ensures that the role of the
subtraction constant is played by the scattering length (see below).
A different choice would have led to the value of the scattering amplitude
at a kinematic point that does not correspond to the threshold to be
the effective subtraction point.  
The fixed-$t$ dispersion relation for $T^+$ is given by
\begin{equation}
\begin{split}
\label{eq:drTplus}
   T^{+}(s,t,u) & = 8 \pi (m+M) a^{+}_0 +
                      \frac{1}{\pi}\!\!\!\!\int_{(m+M)^2}^{\infty}\!\!\!\!
                      \frac{d\,s'}{{s'}^2}\left[\frac{s^2}{s'-s}+\frac{u^2}
                        {s'-u}\right] A^{+}_s(s',t)\\
                &   + S^+ + L^+(t) + U^+(t).
\end{split}
\end{equation}
The expressions for $S^+, L^+(t)$, and $U^+(t)$ can be found in \cite{JN1},
which when adopted to our normalization conventions for the amplitude are
\bean
   S^{+} & = & \frac{1}{2\pi}\!\!\!\int_{(m+M)^2}^{\infty}\!\!\!d\,s'
            \,\,   \frac{\Delta^2-s'\Sigma}
              {q_{s'}^2 {s'}^2} A^{+}_s(s',t'_{\Delta^2}),\\
   L^{+}(t) & = & \frac{t}{\pi}\int_{4 M^2}^{\infty}\frac{d\,t'}
                    {t'(t'-t)} A^{+}_t(t',\Delta^2),\\
   U^{+}(t) & = & \frac{1}{\pi}\!\!\!\int_{(m+M)^2}^{\infty}\!\!\!d\, s'
                  \frac{ s'(2\Sigma-t)-2\Delta^2}{
                      {s'}^2 (4 q_{s'}^2+t)}A^{+}_s(s',t'_{\Delta^2})\\
            &  & - \frac{1}{\pi}\!\!\!\int_{(m+M)^2}^{\infty}\!\!\!d\,s'
                   \frac{s'(2\Sigma-t)^2-2\Delta^2 s' - \Delta^2(2\Sigma-t)}{
                     {s'}^3 (4 q_{s'}^2+t)}A^{+}_s(s',t).
\been
It is important to keep in mind that the absorptive parts
$A_s^+(s',t'_{\Delta^2})$ and $A_t^+(t',\Delta^2)$ are evaluated on
the hyperbola defined by $s'\cdot u'=\Delta^2$. The hyperbolic dispersion
relation for $s$ and $u$ lying on a hyperbola $s\cdot u =b$ may be written as
\begin{equation}\label{eq:drhTplus}
\begin{split}
   T^+(t,b) & = 
   \frac{t}{\pi}\int_{4M^2}^\infty\frac{d\,t'}{t'}
   \frac{A^+_t(t',b)}{t'-t}  
    +\frac{1}{\pi}\!\!\!\!\!\int_{(m+M)^2}^\infty\!\!\!\!\!\frac{d\,s'}{s'}
   \left[\frac{s}{s'-s}+\frac{u}{s'-u}\right] A^+_s(s',t_b') + h(b),
\end{split}
\end{equation}
where the explicit expression for $h(b)$ may be found in ref.~\cite{JN1}.
We do not exhibit it here since this expression only enters the computation
for the $t-$ channel partial wave equation, and does not directly enter our
considerations.
It must also be
noted that combining fixed-$t$
and hyperbolic dispersion relations yields an effective dispersion relation
on which there are no crossing constraints at a fixed value of $s\cdot u =b$.

For the amplitude $T^-(s,t,u)$ we introduce a new dispersion relation.  This 
is achieved by first considering
\begin{equation}\label{eq:drTminus}
\begin{split}
   T^{-}(s,t,u) & = 
          \frac{1}{\pi}\!\!\!\int_{(m+M)^2}^{\infty}\!\!\!\frac{d\,s'}{{s'}^2}
          \left[\frac{s^2}{s'-s}-\frac{u^2}{s'-u}\right] A^-_s(s',t)
                 + d(t) (s-u).
\end{split}
\end{equation}
The subtraction function $d(t)$ is determined by 
writing down a hyperbolic dispersion relation on $s\cdot u = b$ 
for $\tilde{T}^-(t,b) = T^-(t,b)/(s-u)$
\begin{equation}\label{eq:drhTminus}
\begin{split}
   \tilde{T}^-(t,b) & = 
   \frac{t}{\pi}\int_{4M^2}^\infty\frac{d\,t'}{t'}
   \frac{\tilde{A}^-_t(t',b)}{t'-t}  
    +\frac{1}{\pi}\!\!\!\!\!\int_{(m+M)^2}^\infty\!\!\!\!\!\frac{d\,s'}{s'}
   \left[\frac{s}{s'-s}+\frac{u}{s'-u}\right] \tilde{A}^-_s(s',t_b') + g(b).
\end{split}
\end{equation}
We note that these dispersion relations are guaranteed to converge
since the fixed-$t$ dispersion is already twice-subtracted and
a singly subtracted dispersion relation for $\tilde{T}^-$ is equivalent
to a twice-subtracted dispersion relation for $T^-$.
By equating eq. (\ref{eq:drTminus}) and eq. (\ref{eq:drhTminus}) at
$t=0$ and $b=\Delta^2$ we find:
\bean
   d(t) & = & 2 \pi\frac{m+M}{m\, M}\,a^{-}_0 + S^- + L^-(t) +
               U^-(t),\\[2mm]
   S^{-} & = & \frac{1}{2 \pi}\!\!\!\int_{(m+M)^2}^{\infty}\!\!\!d\,s'
               \frac{\Delta^2-s'\Sigma}%
              {s' q_{s'}^2 ({s'}^2-\Delta^2)} A^{-}_s(s',t'_{\Delta^2}),\\
   L^{-}(t) & = & \frac{t}{\pi}\int_{4 M^2}^{\infty}\frac{d\,t'}%
                    {t'(t'-t)} \tilde{A}^{-}_t(t',\Delta^2),\\
   U^{-}(t) & = & \frac{1}{\pi}\!\!\!\int_{(m+M)^2}^{\infty}\!\!\!d\, s'
                  \left[\frac{1}{\Delta^2-{s'}^2}+\frac{1}{s'(4
                      q_{s'}^2+t)}\right] A^{-}_s(s',t'_{\Delta^2})\\
            &  &\qquad +\frac{1}{\pi}\!\!\!\int_{(m+M)^2}^{\infty}\!\!\!
                   d\,s'\left[\frac{1}{{s'}^2}-\frac{1}{s'(4 q_{s'}^2 +
                t)}\right]A^{-}_s(s',t).
\been
The corresponding expression for $g(b)$ may be computed by following
the procedure that led to the expressions above, and is not
exhibited here since this expression only enters the computation
for the $t-$ channel partial wave equation.
\subsection{Dispersion relations with \boldmath{$S-$} and \boldmath{$P$--}wave
absorptive parts}
To perform a comparison of the amplitudes $T^{\pm}$ in their chiral
and dispersive framework, we saturate the above fixed-$t$ dispersion
relations with $S$- and $P$-waves. As it is a straightforward 
calculation, we give here only two examples of contributions from the
dispersion relation for $T^{+}$, showing the interplay between chiral
and dispersive representations. The integral in eq. (\ref{eq:drTplus})
can be written as
\begin{equation}\label{eq:ex1}
\begin{split}
   \frac{1}{\pi}\!\!\!\int_{(m+M)^2}^{\infty}\!\!\!
        \frac{d\, s'}{{s'}^2}\left[\frac{s^2}{s'-s}+
\frac{u^2}{s'-u}\right]A^+_s(s',t)
         = -12\Delta^2\left( s+u\right)\!\!\!\!\!\!
          \int_{(m+M)^2}^{\infty}\!\!\!\!\!\!\frac{d\,s'}{{s'}^3}
           \frac{1}%
           {q_{s'}^2}{\rm Im\,}f^{+}_{1}(s')\hspace*{20mm}&\\
           +16 \left(
             s^2+u^2\right)\!\!\!\!\!\!\int_{(m+M)^2}^{\infty}\!\!\!\!\!\! 
            \frac{d\,s'}{{s'}^3}
            \left[ {\rm Im\,}f^{+}_0(s') + 
              \frac{3{s'}}{4 q_{s'}^2}{\rm Im\,}f^{+}_1(s')\right]
           +16\!\!\!\!\!\!\int_{(m+M)^2}^{\infty}\!\!\!\!\!\!
             \frac{d\,s'}{{s'}^3}\left[\frac{s^3}{s'-s}+
             \frac{u^3}{s'-u}\right]&{\rm Im\,}f^{+}_0(s')\\
           + 12\,s^2\left(t-u+\frac{\Delta^2}{s}\right)\!\!\!\!\!%
            \int_{(m+M)^2}^{\infty}\!\!\!\!\!
            \frac{d\,s'}{{s'}^2}\frac{1}{ q_{s'}^2}
            \frac{{\rm Im\,}f^+_1(s')}{s'-s}
          + 12\,u^2\left(t-s+\frac{\Delta^2}{u} \right)\!\!\!\!\!%
            \int_{(m+M)^2}^{\infty}\!\!\!\!\!
            \frac{d\,s'}{{s'}^2}&\frac{1}{q_{s'}^2}
            \frac{{\rm Im\,}f^+_1(s')}{s'-u}.
\end{split}
\end{equation}
Here one finds polynomials in $s$ and $u$ and integrals which are identical
in structure to three of the integrals in  eq. (\ref{eq:chiral_Tplus}).
The last of the integrals in eq.~(\ref{eq:chiral_Tplus}) has a structure
whose dispersive counterpart arises from $L^+(t)$.  Furthermore,
$U^+(t)$ is quadratic in $t$ in the $S$- and $P$-wave approximation,
\begin{eqnarray*}
   U^+(t) & = & - 24\, t^2 \int_{(m+M)^2}^\infty 
                  \frac{d\, s'}{{s'}^2 q_{s'}^2}{\rm Im\,}f_{1}^{+}(s') \\
        &   & + 16\, t \int_{(m+M)^2}^\infty \frac{d\,s'}{{s'}^2}
               \left( \frac{3({s'}^2+6 \Sigma s'-\Delta^2)
               }{4 s' q_{s'}^2}{\rm Im\,}f_{1}^{+}(s')-{\rm
               Im\,}f^+_0(s') \right)\\ 
        &   & + 32 \int_{(m+M)^2}^\infty \frac{d\,s'}{{s'}^2} \left(
              \Sigma\,{\rm Im\,}f_{0}^{+}(s') + \frac{3\left(\Delta^2
                  \Sigma - s'\left( 
                  2 \left(\Sigma^2-\Delta^2\right)+\Sigma s'\right)\right)}
            {4 s' q_{s'}^2}{\rm Im\,}f^+_1(s')\right). 
\end{eqnarray*}
We take this opportunity to note that the contribution of a state of angular
momentum $l$ to $U^+$ is a polynomial of degree $l+1$, while the contribution
to $U^-$ is a polynomial of degree $l$.  However, there does not
appear to be any elegant closed form expression for such contributions, which
will be of interest in the subsection on contributions from higher waves.
 
Treating the remaining parts of eqs. (\ref{eq:drTplus}) and
(\ref{eq:drTminus}) in a similar way, the polynomial part of $T^{\pm}$
can be written, after eliminating the ambiguity associated with the mass
shell condition $s+t+u=2 \Sigma$ as\footnote{The integrals of the
  dispersive representation of $T^{\pm}$ are identical to the ones in
  eqs.~(\ref{eq:chiral_Tplus}) and (\ref{eq:chiral_Tminus}), respectively.}
\begin{eqnarray}
   T^{+}_{P,disp.}(s,t,u) & = & x_1 + 2 \Sigma x_2 + 2 \Sigma^2 x_3+  
                                (x_4 - 2 \Sigma x_3 -x_2) t \nonumber \\
                    &   &  + \frac{1}{2} t^2
                            (x_3 + 2x_5) +\frac{1}{2} (s-u)^2 x_3,
                            \label{sppoly} \\ 
   T^{-}_{P,disp.}(s,t,u) & = & (s-u)\left\{2\pi\frac{m+M}{m M} a^{-}_{0} + y_1+
                          2 \Sigma y_2 + y_3+y_5\right\} + t(s-u)
                          \left\{y_4-y_2+y_6\right\}, \nonumber
\end{eqnarray}
where
\bea
   x_1 & = & 8\pi (m+M) a^{+}_0 
             +8\!\!\!\!\!\!\int_{(m+M)^2}^{\infty}\!\!\!\!\!\!d\,s'
             \frac{\left( {\Delta}^2 - \Sigma \,s' \right) \,}
             {{s'}^2 q_{s'}^2}\left( {\rm Im\,}f^{+}_{0}(s') -
               3\,{\rm Im\,}f_{1}^{+}(s') \right)\nonumber\\
       &    & + 32\!\!\!\!\!\!\int_{(m+M)^2}^\infty\!\!\!\!%
                \frac{d\,s'}{{s'}^2}\left( \Sigma \,{{\rm
                      Im\,}f_{0}^{+}}(s') + \frac{3\left[
                      {\Delta}^2\,\Sigma  - s'\, \left\{
                        2(\Sigma^2-\Delta^2) + \Sigma\,s' \right\}
                    \right]} 
                    {4 s' q_{s'}^2 }{\rm Im\,}f_{1}^{+}(s') \right),\nonumber\\
   x_2 & = & -12 \Delta^2\!\!\!\!\!\!\int_{(m+M)^2}^{\infty}\!\!\!\!
                           \frac{d \,s'}%
                           {{s'}^3 q_{s'}^2}{\rm Im\,}f^{+}_{1}(s'),\nonumber\\
   x_3 & = & 16 \!\!\!\int_{(m+M)^2}^{\infty}
                           \!\!\!\frac{d\,s'}{{s'}^3}\left[ {\rm
                               Im\,}f^{+}_0(s') + 
                           \frac{3 s'}{4 q_{s'}^2}
                           {\rm Im\,}f^{+}_1(s')\right]\label{eq:x},\\
   x_4 & = & \frac{16}{\sqrt{3}}\int_{4 M^2}^\infty
                               \frac{d\,t'}{{t'}^2}{\rm
                                 Im\,}f^{I_t=0}_0(t')\nonumber\\ 
                    &    &  \phantom{+t\left(\right.}+
                            16\!\!\!\int_{(m+M)^2}^\infty\!\!\!\!%
                            \frac{d\,s'}{{s'}^2}
                            \left( \frac{3\,\left( {s'}^2  + 
                                  6\,\Sigma \,s'
                                 -{\Delta}^2 \right) \,
                                {{\rm Im\,}f_{1}^{+}}(s')}{4 s' q_{s'}^2 } -
                              {{\rm Im\,}f_{0}^{+}}(s') \right),\nonumber\\
   x_5 & = & \frac{16}{\sqrt{3}}\int_{4 M^2}^\infty
                                 \frac{d\,t'}{{t'}^3}{\rm Im\,}f^{I_t=0}_0(t')
                               - 24 \int_{(m+M)^2}^\infty \!\!\!\!\!\!\frac{d\,s'}
                               {{s'}^2 q_{s'}^2}{\rm Im\,}f_{1}^{+}(s').\nonumber
\bee
and
\bea
   y_1 & = & \int_{(m+M)^2}^{\infty}\!\!\!d\,s'\frac{-12\Delta^2}
                  {{s'}^3 q_{s'}^2}{\rm Im\,}f^{-}_1(s'),\nonumber\\
   y_2 & = & \int_{(m+M)^2}^{\infty}\!\!\!\frac{d\,s'}{s'}\left[\frac{16}{{s'}^2}
                  {\rm Im\,}f^{-}_0(s') + 
                  \frac{12}{s' q_{s'}^2}{\rm Im\,}f^{-}_{1}(s')\right],\nonumber\\
   y_3 & = & 8 \!\!\!\int_{(m+M)^2}^{\infty}\!\!\!d\,s'
                  \frac{\Delta^2-\Sigma s'}%
                  {s' q_{s'}^2 ({s'}^2-\Delta^2)}\left({\rm Im\,}f^{-}_0(s') -
                  3 {\rm Im\,}f^{-}_1(s')\right),\nonumber\\
   y_4 & = & 6\sqrt{2} \int_{4 M^2}^{\infty}\frac{d\,t'}{{t'}^2}
                 {\rm Im\,}\frac{f^{I_t=1}_1(t')}{p_{t'} q_{t'}},\label{eq:y}\\
   y_5 & = & \int_{(m+M)^2}^{\infty}\!\!\!
                  d\,s'\frac{16\,\Delta^2}
                  {{s'}^2\,\left( \Delta^2 - {s'}^2 \right) }
                  {\rm Im\,}f^{-}_{0}(s')\nonumber\\
     &   & + \!\!\!\!\!\!\int_{(m+M)^2}^{\infty}\!\!\!\!\!\!
             d\,s'\frac{12\,\left( {\Delta}^4
           - 2\,{\Delta}^2\,\Sigma \,s' -
          3\,{\Delta}^2\,{s'}^2 + 4\,\Sigma \,{s'}^3
          \right)}{{s'}^3  q_{s'}^2 \,
         \left( {\Delta}^2 - {s'}^2 \right) }{\rm Im\,}f^{-}_{1}(s'),\nonumber\\
   y_6 & = & \!\!\!\!\!\!\int_{(m+M)^2}^{\infty}\!\!\!\!\!\!
           d\,s'\frac{24}{{s'}^2 q_{s'}^2}{\rm
                  Im\,}f^{-}_{1}(s')\nonumber .
\bee
In the manner described above, we have established the starting point for
the comparison of the contributions of $S$-- and $P$--wave absorptive parts
to the low energy polynomial.  

\subsection{Contributions from higher partial waves}
Contributions from higher partial waves are twofold. On the
one hand they contribute to polynomials as in eq.~(\ref{sppoly}). On
the other hand these waves also yield additional dispersive integrals
similar to the last three terms in eq.~(\ref{eq:ex1}). However,
applying the chiral power counting scheme, one can see that the
corresponding chiral integrals are of order $O(q^6)$, so that they are
neglected in chiral perturbation theory to one loop. Therefore, only the contributions of
the higher partial waves to the polynomials in eq.~(\ref{sppoly}) are
of interest here.
It may be readily seen that the $l \geq 2$ partial waves contribute
to the low energy polynomial of $T^{+}$at this level as follows:
the contribution coming from the fixed-$t$ dispersive integral of
eq.~(\ref{eq:drTplus}) to the coefficient of $(s^2+u^2)$ reads
\begin{eqnarray*}
  \zeta_{ft}& = & 16 \sum_{l=2}^{\infty} (2 l + 1)\!\!\!\int_{(m+M)^2}^{\infty}
                    \!\!\!\frac{d\, s'}{{s'}^3}{\rm Im\,}f^+_l(s').
\end{eqnarray*}
The contributions to the coefficient of $t$ and $t^2$ of the polynomial coming
from $L^+(t)$ read
\bean
  \zeta_{{L}_1}       & = & \frac{16}{\sqrt{3}} \sum_{l=2}^{\infty} (2 l + 1)
                            \int_{4 M^2}^\infty
                            \frac{d\,t'}{{t'}^2}{\rm Im\,}f^{I_t=0}_l(t'),\\
  \zeta_{{L}_2}       & = & \frac{16}{\sqrt{3}}    \sum_{l=2}^{\infty} (2 l + 1)
                            \int_{4 M^2}^\infty
                            \frac{d\,t'}{{t'}^3}{\rm Im\,}f^{I_t=0}_l(t'),
\been
whereas $S^+$ contributes to the constant part of
the low energy polynomial,
\bean
   \zeta_S & = & 8\sum_{l=2}^{\infty}(-1)^l (2 l + 1)
                 \!\!\!\!\!\!\int_{(m+M)^2}^{\infty}\!\!\!\!\!\!d\,s'
                 \frac{\left( {\Delta }^2 - \Sigma \,s' \right) \,
                   }{{s'}^2 q_{s'}^2 }{\rm Im\,}f^{+}_l (s').
\been
As noted earlier, the contribution of a partial wave of angular
momentum $l$ to $U^{+}$ is a polynomial in $t$ of degree $l+1$, and the
three lowest coefficients contributing to eq.~(\ref{sppoly}) can be
read of from it. Then the expression below is the sum of all such
contributions: 
\begin{eqnarray*}
  \zeta_{{U}_0}+\zeta_{{U}_1}t+\zeta_{{U}_2}t^2.
\end{eqnarray*}

An analogous procedure for the contributions to $T^-$ from the
higher waves may also be performed.  The contributions from the
fixed-$t$ dispersive integral of eq.~(\ref{eq:drTminus}) will make a
contribution proportional to $(s^2-u^2)$ whose coefficient is
\begin{eqnarray*}
  \xi_{ft}& = & 16 \sum_{l=2}^{\infty} (2 l + 1)\!\!\!\int_{(m+M)^2}^{\infty}
                    \!\!\!\frac{d\, s'}{{s'}^3}{\rm Im\,}f^-_l(s').
\end{eqnarray*}
There is a contribution coming from $L^-(t)$ proportional
to $t$ which reads 
\begin{eqnarray*}  
   \xi_{L}       & = & 2\sqrt{2} \sum_{l=3}^{\infty} (2 l + 1)\int_{4 M^2}^\infty
                       d\,t'\frac{1}{{t'}^2}{\rm
                       Im\,}\frac{f^{I_t=1}_l(t')}
                     {q_{t'}p_{t'}}, 
\end{eqnarray*}
and a contribution from $S^-$ which reads
\bean
   \xi_S & = & 8\sum_{l=3}^{\infty} (-1)^l(2 l + 1)\!\!\!\!\!\!\int_{(m+M)^2}^{\infty}
               \!\!\!\!\!\!d\,s'\frac{\left( {\Delta }^2 - 
                   \Sigma \,s' \right) \,{\rm Im\,}f^{-}_l  (s')}
               {{s'}^2 q_{s'}^2(s'-\Delta^2)}.
\been
Here we merely denote
the sum of such contributions from $U^-$ to the low energy polynomial by
\begin{eqnarray*}
  \xi_{U^0}+ \xi_{U^1} t 
\end{eqnarray*}
The resulting polynomials then read
\begin{eqnarray}
T_{hw}^+(s,t,u) & = &2 \Sigma^2 \zeta_{ft} + \zeta_S + \zeta_{{U}_0} + (\zeta_{{L}_1}+
                  \zeta_{{U}_1}-2 \Sigma \zeta_{ft})t \nonumber \\ 
           &   & + (\zeta_{{L}_2}+\zeta_{{U}_2}  
                 + \frac{\zeta_{ft}}{2})t^2 + \frac{\zeta_{ft}}{2} (s-u)^2, 
\label{higherpoly} \\
T_{hw}^-(s,t,u)& =& (s-u)(2\, \Sigma\, \xi_{ft}+ \xi_S + \xi_{U^0} +(\xi_{L} +
                    \xi_{U^1}-\xi_{ft})t). \nonumber 
\end{eqnarray}
In summary, the dispersive representation for the low energy polynomial
is the sum of the contributions arising from the $S$-- and $P$--wave absorptive
parts, eq.~(\ref{sppoly}), and those of the higher partial waves,
eq.~(\ref{higherpoly}). 
Once the dispersive representation is saturated with phenomenological
absorptive parts and is compared with the chiral representation
eq.~(\ref{chiralpoly}), then the procedure would amount to a
determination of the low energy  constants of chiral perturbation theory. 
\subsection{Partial wave equations}
Analyticity, unitarity, and crossing symmetry lead to a set of
integral equations (Steiner-Roy equations) relating each of the
partial waves to all the other ones \cite{Lang,Steiner,Roy}. These
equations depend on the choice of dispersion relations. 
In the case at hand, the  integral equations are derived by projecting
eqs.~(\ref{eq:drTplus})-(\ref{eq:drhTminus})
onto partial waves and by inserting a partial wave expansion for the
absorptive parts.   For Steiner--Roy equations based on other dispersion
relations, see
ref.~\cite{NielsenOades,Hedegaard-Jensen,JohannessonandPetersen}.
In contrast to the dispersion relations for the full amplitudes,
the range of validity of these equations is restricted by the Lehmann
ellipse(s).   Assuming Mandelstam analyticity, the sum of partial waves
for $A^{\pm}(s,t)$ is convergent for all $s$ if $-32 M^2 \leq t \leq
4 M^2$, implying that the partial wave equations for the $s$-channel
waves are valid in the range $2.43 M^2 \leq s \leq 57.14 M^2$. 
Analogously, for the $t$-channel partial wave equations, the range
of validity is $-28.2 M^2\leq t \leq 82.2 M^2$ \cite{Lang}.

The Steiner--Roy equations for the $s$-channel $S$- and $P$-waves from
$T^+$ read (in the $S$- and $P$-wave approximation)
\begin{align}
\begin{split}\label{eq:roy_plus}
   f^+_l(s)  =  \delta_{0,l}\, a^+_0\,\frac{m+M}{2}  +
        \!\!\!\!\!\!\int_{(m+M)^2}^\infty\!\!\!\!\!\! d\,s'
        K^+_{l,0}(s,s') {\rm Im}\, f^+_0(s')&\\
      + \!\!\!\!\!\!\int_{(m+M)^2}^\infty\!\!\!\!\!\! d\,s' K^+_{l,1}(s,s')
        {\rm Im}\, f^+_1(s') 
      +\!\!\! \int_{4 M^2}^\infty\!\!\!& d\,t' K^{(0)}_{l,0}(s,t') {\rm Im}\,
        f^{I_t=0}_0(t'),\\
   f^{I_t=0}_{0}(t)  =  \frac{\sqrt{3}}{2}(m+M) a^+_0
   +\!\!\!\!\!\!\int_{(m+M)^2}^\infty\!\!\!\!\!\! d\,s'
          G^{+}_{0,0}(t,s') {\rm Im}\, f^+_0(&s')\\
        +\!\!\!\!\!\!\int_{(m+M)^2}^\infty\!\!\!\!\!\! d\,s'
         G^{+}_{0,1}(t,s'){\rm Im}\,f^+_1(s')
         +\!\!\!\int_{4 M^2}^\infty\!\!\!& d\,t'
         G^{(0)}_{0,0}(t,t'){\rm Im}\, f^{I_t=0}_0(t'),
\end{split}
\end{align}
while the ones obtained from $T^-$ are
\begin{align}
\begin{split}\label{eq:roy_minus}
   f^{-}_l(s) & =  \delta_{l,0}\, a^{-}_0\,\frac{(m+M)}{2} \,\frac{3 s^2 -
              2 s (m^2+M^2)-(m^2-M^2)^2}{8 \,s\, m\, M}\\
              &\hspace*{20mm} +\delta_{l,1} a^{-}_0\frac{(m+M)}{2} \,\frac{m^4+(M^2-s)^2-2
              m^2(M^2+s)}{24\,s\,m\,M}\\
              & +
        \!\!\!\!\!\!\int_{(m+M)^2}^\infty\!\!\!\!\!\! d\,s'
        K^-_{l,0}(s,s') {\rm Im}\, f^-_0(s')
      + \!\!\!\!\!\!\int_{(m+M)^2}^\infty\!\!\!\!\!\! d\,s' K^-_{l,1}(s,s')
        {\rm Im}\, f^-_1(s') 
      +\!\!\! \int_{4 M^2}^\infty\!\!\! d\,t' K^{(1)}_{l,1}(s,t') {\rm Im}\,
        \frac{f^{I_t=1}_1(t')}{q_{t'}p_{t'}},\\
   f^{I_t=1}_{1}(t) & =  a^{-}_{0}\,
   \frac{(m+M)}{2}\,\frac{\sqrt{t-4m^2}\sqrt{t-4M^2}}{6\sqrt{2} m\,M}\\
                    & +\!\!\!\!\!\!\int_{(m+M)^2}^\infty\!\!\!\!\!\! d\,s'
          G^{-}_{1,0}(t,s') {\rm Im}\, f^-_0(s')
        +\!\!\!\!\!\!\int_{(m+M)^2}^\infty\!\!\!\!\!\! d\,s'
         G^{-}_{1,1}(t,s'){\rm Im}\,f^-_1(s')
         +\!\!\!\int_{4 M^2}^\infty\!\!\! d\,t'
         G^{(1)}_{1,1}(t,t'){\rm Im}\, \frac{f^{I_t=1}_1(t')}{q_{t'}p_{t'}}.
\end{split}
\end{align}
The kernel functions $K, G$ can be found in appendix C.

We note here that this is effectively a system of closed equations for
the $S$-- and $P$--waves.  In order to solve them in the low-energy region,
in practice the contributions of the absorptive
parts of the $l\geq 2$ waves and that of the high energy tail of the
$S$-- and $P$--waves are added together to yield the driving terms for
this system  from the dispersion relations
eqs.~(\ref{eq:drTplus})-(\ref{eq:drhTminus}),
when expressions for the absorptive parts of the $l\geq 2$ are
inserted into the right hand sides and by writing down forms for the $S$--
and $P$--waves compatible with unitarity and with the requirement that
they reproduce the scattering lengths. Such a program has been
recently carried out for $\pi\pi$ scattering, see ref.~\cite{acgl}.
\section{Low Energy Constants from Phenomenology}
The coefficients of the chiral polynomials, eq.
(\ref{chiralpoly}), are functions of the low energy constants $L^r_i$
whereas the coefficients
of the dispersive polynomial part of the amplitudes, eq.
(\ref{sppoly}), are given in terms of integrals over the lowest six
partial waves.
Once the imaginary parts of these are known, a comparison with the
chiral polynomials yields 
the low energy constants involved in $\pi K$ scattering. As 
eqs. (\ref{chiralpoly}) and (\ref{sppoly}) only provide six equations,
only constraints for combinations of the seven low energy constants involved
can be derived. 

In the present work we focus  on the combinations $4L_2^r + L_3$, $4
L_1^r + L_3 - 4 L_4^r - L^r_5 + 4 L_6^r + 2 L_8^r$, and $F_\pi F_K + 4
L^r_5 (M^2 +m^2)$. The first appears in the term proportional to
$(s-u)^2$ in $T^+$, while the second  one comes from the constant part
of the same amplitude. The last combination stems from the
$t$-independent part of the amplitude $T^-$. Only $4L_2^r + L_3$ does
not explicitly depend on the scattering lengths. Therefore 
we expect
this combination to be the one which can be estimated most precisely.

To evaluate the coefficients of the dispersive polynomials,
eqs.~(\ref{eq:x},\ref{eq:y}), we employ a $K$--matrix
parametrization similar to the ones in \cite{Oller}, but with few
more free parameters, and require the resulting phase shifts to fit the
experimental data of \cite{Estabrooks:1978xe} in the elastic region.
As the integral cut--offs in eqs. (\ref{eq:x}) and (\ref{eq:y}) we choose the
elasticity thresholds $1.69\mbox{ GeV}^2$ for $I=1/2$ and $2.96 \mbox{
  GeV}^2$ for $I=3/2$. Here, we do not take into account the
contributions from the higher partial waves. The masses of the pion,
kaon, and eta, the latter entering the calculation only through the
loop--functions, are set to $M=139.56$ MeV, $m=497.67$ MeV, and
$m_\eta=547.30$ MeV, respectively. The decay constants are
$F_\pi = 92.4$ MeV, $F_K = 1.22 F_\pi$ (we take the well-established
analysis for the ratio $F_K / F_\pi$ in the present work; new analyses
are now available \cite{Fuchs}, and these will be incorporated at the time the
fresh Steiner-Roy equation fits to the data are ready \cite{BM}). 
Furthermore, the renormalization scale $\mu$ is set to $m_\rho
=769.30$~MeV. For the three combinations of low energy constants we obtain 
\bea
   4 L_2^r + L^r_3 & = & 0.0027\pm 0.0001 ,\nonumber\\
   4 L_1^r + L^r_3 - 4 L_4^r - L^r_5 + 4 L_6^r + 2 L_8^r & = & -0.0003\pm
   0.0013+ 0.14\, {\rm GeV}\cdot a^{+}_0, \label{estimates1}\\
   L^r_5 & = & -0.0065 \pm 0.0001 + 0.024\, {\rm GeV}\cdot a^-_0,\nonumber
\bee
where $a^\pm_0$ are given in GeV$^{-1}$. The quoted errors are due to
varying the integral cut--offs by $20 \%$. Note that the coefficients
of the scattering lengths, i.e. $0.14\,\mbox{GeV}^{-1}$ and
$0.024\,\mbox{GeV}^{-1}$, are fixed by chiral perturbation theory.

In order to check the influence of the parametrization on the above
numerical results, we have chosen yet another $K$--matrix
parametrization with fewer parameters to fit the experimental
data. The quality of these fits is not as good, especially for the
$I=1/2$ $S$--wave and the $I=3/2$ $P$--wave, whereas the other two
waves are not changed significantly. However, as the phases of
the $I=3/2$ $P$--wave are small at low energies, changes to this
partial wave are not important. This parametrization yields
\bea
   4 L_2^r + L^r_3 & = & 0.0029\pm 0.0001 ,\nonumber\\
   4 L_1^r + L^r_3 - 4 L_4^r - L^r_5 + 4 L_6^r + 2 L_8^r & = & -0.012\pm
   0.001+ 0.14\, {\rm GeV}\cdot a^{+}_0, \label{estimates2}\\
   L^r_5 & = & -0.012 \pm 0.00001 + 0.024\, {\rm GeV}\cdot a^-_0,\nonumber
\bee
where again the quoted errors are due to changes in the integration
cut--offs. The numerical coefficients on the right hand sides of
eqs.~(\ref{estimates1},\ref{estimates2}) generally depend on the
renormalization scale $\mu$, as well. Note that the last lines in
eqs.~(\ref{estimates1},\ref{estimates2}) amount to an alternative
method to fix $L^r_5$, which could then be employed to determine the
ratio of the pion and the kaon decay constants. However, as the
predictions of the decay constants are much more reliable than the one
for the $\pi K$ scattering length, we regard the relation between
$L^r_5$ and $a^-_0$ as a consistency check for our method than a
accurate way of determining $L^r_5$.
Comparing the above with the values in \cite{BEG},
\bea
   4 L_2^r + L_3 & = & 0.0019\pm 0.0013,\nonumber\\
   4 L_1^r + L_3 - 4 L_4^r - L^r_5 + 4 L_6^r + 2 L_8^r & = &
   -0.0011\pm 0.0018 
   \label{estimates3}\\
   L^r_5 & = & 0.0014\pm 0.0005,\nonumber
\bee
one can see that our values for the first combination are in reasonable
agreement with the previous determination. The values for the second
combination of low energy constants are still reasonable if the wide
spread of experimental values for the scattering lengths (see \cite{BKM}
and references therein),
\bean
   -0.31\, {\rm GeV}^{-1}\leq a^{+}_0 \leq 0.34\, {\rm GeV}^{-1}, \quad
    0.43\, {\rm GeV}^{-1}\leq a^{-}_0 \leq 0.89\, {\rm GeV}^{-1},
\been
is taken into account, whereas the determination of $L^r_5$ is
less reliable, emphasizing the need of a detailed analysis of the $\pi
K$ partial waves and the importance of the contributions of the
higher partial waves to the low energy polynomials.
This will be done elsewhere \cite{BM}.  Furthermore,
there are other recent determinations of the low energy constants
\cite{amoros} and a detailed comparison will be made when the Steiner-Roy
equation fits are available \cite{BM}. 

Comparing the results in eqs.~(\ref{estimates1},\ref{estimates2}), one
can see that the numerical value for the first of the above
combinations for the LECs does not depend very much on the
parametrization, whereas this dependence for the second and the third
combination is more substantial. Note, however, that this dependence
is accommodated by the wide spread of experimental values for the
scattering lengths, again calling for a more detailed analysis of the
phase shifts and the scattering lengths.
\section{Summary and Conclusions}
The $\pi K$ scattering problem is an important process in the
low energy sector of the strong interactions.  Compared to the
closely related problem of $\pi \pi$ scattering, considerably less
is known about this process due to the lack of availability of
experimental data, the relative paucity of theoretical results
associated with the absence of three-channel crossing symmetry, 
and the presence of unequal masses of the particles.
Despite these difficulties, here we have shown that the results of
the type established in the recent past for $\pi \pi$ scattering
can be extended to the $\pi K$ case.  We have noted that the
experimentally accessible scattering length $a^-_0$ at one-loop
order in chiral perturbation theory is essentially parameter free
and that a precise determination of this quantity would constitute
a precision test of chiral perturbation theory.

We have established a framework within which the $\pi K$ amplitude in $SU(3)$
chiral perturbation theory can be split up into functions of single variables
and are then replaced by an dispersive representation that leads to an
effective low energy polynomial representation and a dispersive tail.
We have considered twice--subtracted fixed-$t$ dispersion relations 
with the subtraction functions determined in terms of dispersion relations
on hyperbolic curves, in particular those for $T^+$ which were
established sometime ago, and new ones for $T^-$.
This allows us to generate a low energy polynomial and
a dispersive tail with the same structure as the chiral representation.  
We have also discussed in some detail the contributions of the
absorptive parts of the $l\geq 2$ partial waves.

Furthermore, explicit integral equations for
the $S$-- and $P$--waves are given which form a closed system when the $l\geq 2$
wave absorptive parts are neglected.  The contributions from those
waves and the high energy part of the $S$-- and $P$--wave absorptive parts
would then determine the driving terms for these (Steiner-Roy) integral
equations.  In particular, a detailed fit of experimental information
to these equations would lead to a precise determination of $a^-_0$.

The comparison of chiral perturbation and dispersion theory representation
of the amplitudes yields a system of sum rules for low energy constants of
chiral perturbation theory.  We have used the results from a recent study
of the phase shift and elasticity information to evaluate certain combinations
of coupling constants, which do not involve the contributions coming from
the t-channel absorptive parts.
First estimates for the $SU(3)$ low energy constants obtained from 
this phenomenology yield the estimates, see 
eq.~(\ref{estimates1}). 
These estimates compare favorably with the determinations reported in
the literature.  A full partial wave analysis will lead to an accurate
evaluation of the coupling constants of interest.  Such a partial wave
analysis combined with chiral inputs can produce reliable estimates for
$\pi K$ scattering lengths which can in principle be measured at pion-kaon
atom ``factories'' such as DIRAC.

\subsection*{Acknowledgements:}  We are particularly indebted to
H.~Leutwyler and U.-G.~Mei{\ss}ner for discussions and comments on the
manuscript.  We also thank B.~Moussallam, J.~A.~Oller, J.~P.~Pel{\'a}ez,
and J.~Stahov for discussions and comments.
B.~A. thanks the Institut f\"ur Kernphysik, Forschungszentrum
J\"ulich, for its hospitality when part of this work was done.
P.~B. thanks the Centre for Theoretical Studies, Indian Institute of
Science, Bangalore, for its hospitality when this work was initiated.
\begin{appendix}
\section{\boldmath{$\pi\pi$} amplitude in \boldmath{$SU(3)$} chiral
  perturbation theory} 
In this brief appendix, we consider the $\pi\pi$ scattering amplitude
presented in ref.~\cite{BKM,dobado}.  It is possible to split
the amplitude $A(s,t,u)$ into three functions of one variable
$W_i(s), i=1,2,3$,
whose absorptive parts may be expressed in terms of those of the three
lowest partial waves $f^I_0, I=0,2$ and $f^1_1$.
In terms of these functions, we may write $A(s,t,u)$ as
\begin{eqnarray}
   A(s,t,u) & = & 32 \pi \left\{\frac{1}{3}W_0(s)+\frac{3}{2}(s-u) W_1(t)+
\frac{3}{2}(s-t) W_1(u) + \frac{1}{ 2} \left(W_2(t) + W_2(u) -
  \frac{2}{ 3} W_2(s) \right) \right\}. \nonumber
\end{eqnarray}
We list one choice of for the functions $W_i, \, i=0,1,2$ in appendix B.
One may now write the $W_i$ in terms of dispersion relations and
generate a low energy polynomial 
representation.  As an illustration we use the dispersive polynomial
established in ref.~\cite{AB1} to evaluate the $SU(3)$ low energy constants
with the three sets of phase shifts described there.  These results are
presented in Table 1 (masses, decay constants and renormalization scale as
in sec. 5).

\begin{table}
$$
\begin{tabular}{||c| c| c| c||}\hline
 & Set 1 & Set 2 & Set 3              \\ \hline
$10^3 L_2^r $ & $1.63$ & $1.63$ & $1.62$     \\ \hline
$ 10^3(2 L_1^r + L_3)$ & $-2.34$ & $-2.28$ & $-2.22$ \\ \hline
$ 10^3(2 L_4^r + L^r_5)$ & $-1.92$ & $-1.56$ & $-1.23$ \\ \hline
\end{tabular}
$$
\caption{$SU(3)$ coupling constants from $\pi\pi$ phase shifts
($\mu=m_\rho$).}
\end{table}

We note that these phase shifts were used to determine the
values for the low energy constants $\overline{l}_i,\, i=1,2,4$ of
$SU(2)$ chiral perturbation theory.
It is well known that when the SU(3) theory is reduced to SU(2)
theory, relations emerge between the low energy constants of the two
theories.  The results for the $\overline{l}_i,\, i=1,2,4$ of ref.~\cite{AB1}
may then be translated into the SU(3) coupling constants which are
listed in Table 2.  Although the results of Table 1 and Table 2
are in general agreement, those in Table 1 amount to a consistent
new determination.  The numbers in Table 1 agree well with determinations
in the literature, see, e.g., ref.\cite{GLNPB,BEG}.  
\begin{table}
$$
\begin{tabular}{||c| c| c| c||}\hline
 & Set 1 & Set 2 & Set 3              \\ \hline
$10^3 L_2^r $ & $1.42$ & $1.41$ & $1.41$     \\ \hline
$ 10^3(2 L_1^r + L_3)$ & $-2.16$ & $-2.10$ & $-2.04$ \\ \hline
$ 10^3(2 L_4^r + L^r_5)$ & $-1.55$ & $-1.20$ & $-0.88$ \\ \hline
\end{tabular}
$$
\caption{$SU(3)$ coupling constants derived from $SU(2)$ effective theory 
($\mu=m_\rho$).}
\end{table}

The phase shift determination of ref.~\cite{AB1} were based on a
Roy equation fit whose driving terms were computed from higher wave
and asymptotic contributions that arose from the $f_2(1270)$ 
and Pomeron and Regge contributions setting in above an energy of 
$\sim 1.5$ GeV, recently described in ref.~\cite{AB2}.
These absorptive parts also contribute to the low energy dispersive
polynomials.  We evaluate the resulting contributions to the low
energy constants whose contributions to $\overline{l}_1$ and $\overline{l}_2$ are 
$\sim -0.1$ and $0.41$, respectively and to $10^3 \, L_2^r$ and 
$10^3(2 L_1^r + L_3)$ are $\sim 0.21$ and $-0.05$ respectively.
We note here that once the scattering length $a^0_0$ is experimentally
determined to within small uncertainties, the Roy equation fits of
ref.~\cite{acgl} may be used to produce sharp values for the combinations
of low energy constants discussed here.

\section{List of functions of single variables}
The functions of single variables entering $T^+(s,t,u)$ are given as
\bean
 16 F_\pi^2 F_K^2  Z^{+}_0(s) & = & \frac{3\, \Delta^2(
L_{K \eta}(s)+ L_{\pi K}(s))}{s} \\
                          &   &  + 2\, \Delta\Sigma(K_{K\eta}(s) + 
                                 K_{\pi K}(s)) + \Sigma^2( 
                                 \frac{J^r_{K \eta}(s)}{3} + 
                                 3\,J^r_{\pi K}(s)) \\
                          &   &  - s\,\left( 4\,F_K\,F_{\pi}+
                                 \Delta(3\,K_{K \eta}(s)\,
                                 + 5\,K_{\pi K}(s))\right. \\
                          &   & \left. \phantom{s\,\left(\right.}+ 
                                16 \Sigma( 8\,L^r_2+2\,L_3+L^r_5)
                           +\Sigma(
                                J^r_{K \eta}(s)+7\,J^r_{\pi K}(s))
                                 \right)\\
                          &   & + s^2\,\left( 64\,L^r_2 + 16\,L_3 + 
                                  \frac{3\,J^r_{K \eta}(s)}{4} + 
                                  \frac{19\,J^r_{\pi K}(s)}{4} \right),\\[2mm]
  16 F_\pi^2 F_K^2 Z^{+}_1(s) & = & -3\,\left( L_{K\eta}(s) + L_{\pi K}(s) - 
                                s\,\left( M^r_{K\eta}(s) + 
                                  M^r_{\pi K}(s) \right)  \right),\\[2mm]
   16 F_\pi^2 F_K^2 Z^{+}_t(s) & = &  
                                      8\,F_K\,F_\pi\,\Sigma + 128 M^2 m^2(4\,L^r_1+L_3 - 
                                      4\,L^r_4  - L^r_5 + 4\,{L_6} + 2\,{L_8})\\
                                      &  & - \frac{16 M^2 m^2 J^r_{\eta \eta }(s)} {9} + 
                                      F_K\,F_\pi\Delta(  3\, \mu_\pi -2\,\mu_{K}  
                                      - \mu_{\eta})\\
                               &   &  + 32 \Sigma^2(4\,{L^r_2} + L_3 + L^r_5)\\
                               &   & - s\,\left(64 \Sigma( 4\, L^r_1 + L_3 - 2\, L^r_4)+ 
                                        2\, M^2(J^r_{\pi \pi}(s) - J^r_{\eta \eta}(s))  
                                      \right)\\
                               &   & + s^2\,\left( 32(4\,L^r_1 + L_3) + 
                                 3\,J^r_{K K}(s) + 
                                 4\,J^r_{\pi \pi }(s) \right),
\been
whereas the functions of single variables entering $T^-(s,t,u)$ read
\bean
96 F_K^2 F_{\pi}^2 Z^{-}_0(s) & = &  
  \frac{18\,\Delta^2(L_{K \eta}(s) + 
     L_{\pi K}(s))}{s}\\
   &   & +  12\, \Delta\Sigma(K_{K \eta}(s)
    + K_{\pi K}(s))+ 
  2\,\Sigma^2(J^r_{K \eta}(s) - 
  3\,J^r_{\pi K}(s))\\
&   & + s\,\left( 24\,F_K\,F_\pi -6 \Delta( 
     3\,K_{K \eta}(s) + 
     5\,K_{\pi K}(s)) - 96 \Sigma( 
     2\,L_3 - 
     \,L^r_5)\right.\\
&   & \left.- 6\Sigma( J^r_{K \eta}(s)-J^r_{\pi K}(s) ) \right)  + s^2\,\left( 96\,L_3 + 
     \frac{9}{2}(J^r_{K \eta}(s) + 
     J^r_{\pi K}(s)) \right),\\[2mm]
96 F_K^2 F_{\pi}^2 Z^{-}_1(s) & = & -18\,(s\,M^r_{K\eta}(s) + s\,M^r_{\pi K}(s) -
                                    L_{K\eta}(s) - L_{\pi K}(s) ),\\[2mm]
96 F_K^2 F_{\pi}^2 Z^{-}_t(s) & = & 24\,s\,(M^r_{KK}(s) + 2 M^r_{\pi\pi}(s) ).
\been
Finally, the functions of single variables required to define the
$\pi\pi$ amplitude in $SU(3)$ chiral perturbation theory can be written as
\begin{eqnarray*}
   W_0(s) & = & \frac{3}{32 \pi}
   \bigg\{\frac{s-M^2}{F_\pi^2}+\frac{1}{F_\pi^4}\left(\frac{M^4}{ 18}
       J^r_{\eta\eta}(s) +\frac{1}{2}(s^2-M^4)^2 J_{\pi\pi}^r(s) +
                \frac{s^2}{8} J^r_{KK}(s) \right) \\
          &   & + \frac{4}{F_\pi^4} \left[ (2 L_1^r+L_3) (s-2 M^2)^2 +
                (4 L_4^r + 2 L_5^r)\, (s-2 M^2)\,M^2 \right. \\
          &   & +\left. (8 L_6^r+4 L_8^r) M^4 \right] \bigg\}+
                W_2(s) \\[2mm]
   W_1(s) & = & \frac{s}{48 \pi} \left\{M^r_{\pi\pi}(s)+\frac{1}{2} M^r_{KK}(s)\right\}
                \\[2mm]
   W_2(s) & = & \frac{1}{16 \pi} \left\{\frac{1}{4 F_\pi^4}(s-2 M^2)^2 J^r_{\pi\pi}(s) +
                \frac{4 L^r_2}{F_\pi^4} (s-2 M^2)^2 \right\}
\end{eqnarray*}
For the definitions of the standard loop functions $J^r_{PQ}(s), M^r_{PQ}(s),
L_{PQ}(s), K_{PQ}(s)$ and $\mu_P$, see ref.~\cite{GLNPB}.   
\section{Kernels of the partial wave equations}
The kernels of the partial wave equations,
eqs.~(\ref{eq:roy_plus},\ref{eq:roy_minus}), are:
\allowdisplaybreaks{
\begin{align*}
   K^+_{0,0}(s,s') & =  \frac{{s'}-2\,s}{\pi \,\left( s - {s'}
     \right) \,{s'}} +  \frac{\ln (s - 2\,{\Sigma} + {s'}) - \ln (s
     - 2\,{\Sigma} + {s'} - 4\,{q_{s}}^2)} {4\,\pi \,{q_{s}}^2} +
   \frac{{{\Delta}}^2 - {\Sigma}\,{s'}} {2\,\pi
     \,{{s'}}^2\,{q_{s'}}^2},\\[3mm]
   K^+_{01}(s,s') & =  \frac{-3\,\left( \ln (s - 2\,{\Sigma} + {s'}) - 
       \ln (s - 2\,{\Sigma} + {s'} - 4\,{q_{s}}^2) \right) \,
     \left( s - 2\,{\Sigma} + {s'} - 2\,{q_{s'}}^2 \right) }{8\,\pi
     \,{q_{s}}^2\,{q_{s'}}^2}\\
     &\quad + \frac{3\,{q_{s}}^2\,\left\{ {{\Delta}}^2\,\left( s - {s'} \right)- 
       {s'}\,\left( 2\,s\,{\Sigma} - 3\,s\,{s'} - 2\,{\Sigma}\,{s'} + 
          {{s'}}^2 \right)  + 4\,{s'}\,\left( s' - s \right)
        \,{q_{s'}}^2 \right\} }{2\,\pi \,\left( s - {s'} \right)
      \,{{s'}}^3\,{q_{s'}}^2}\\
    &\qquad  + \frac{3\,\left\{ \left( s - {s'}
        \right) \, \left( {s'}^3 + {\Delta}^2\,\left( {\Sigma} + {s'}
          \right)-2\,{{\Sigma}}^2\,{s'}  \right)  + 
       2\,{s'}\,\left[ {s'}\,\left( 2\,{\Sigma} + {s'} \right) -
         2\,s\,\left( {\Sigma} + {s'} \right) \right]
        \,{q_{s'}}^2 \right\} } {2\,\pi \,\left( s - {s'} \right)
      \,{{s'}}^3\,{q_{s'}}^2},\\[3mm]
   K^{(0)}_{0,0}(s,t') & =  \frac{ {t'}\,\left\{\ln ({t'} +
         4\,{q_{s}}^2)- \ln ({t'}) \right\}  - 4\,{q_{s}}^2 }
   {4\,{\sqrt{3}}\,\pi \,{t'}\,{q_{s}}^2},\\[3mm]
   K^{+}_{1,0}(s,s') & = \frac{1}{2\,\pi \,{q_{s}}^2} - \frac{\left\{
       \ln (s - 2\,{\Sigma} + {s'}) - 
       \ln (s - 2\,{\Sigma} + {s'} - 4\,{q_{s}}^2) \right\} \,
     \left( s - 2\,{\Sigma} + {s'} - 2\,{q_{s}}^2 \right) }{8\,\pi
     \,{q_{s}}^4},\\[3mm]
   K^{+}_{1,1}(s,s') & =  \frac{-3\,\left( s - 2\,{\Sigma} + {s'} -
       2\,{q_{s'}}^2 \right) } {4\,\pi \,{q_{s}}^2\,{q_{s'}}^2}\\
   &\quad + \frac{3\,\left\{ \ln (s - 2\,{\Sigma} + {s'}) - 
       \ln (s - 2\,{\Sigma} + {s'} - 4\,{q_{s}}^2) \right\} \,
     \left( s - 2\,{\Sigma} + {s'} - 2\,{q_{s}}^2 \right) \,
     \left( s - 2\,{\Sigma} + {s'} - 2\,{q_{s'}}^2 \right) }{16\,\pi
     \,{q_{s}}^4\, {q_{s'}}^2}\\
   &\qquad - \frac{{q_{s}}^2\,\left(
       {{\Delta}}^2\,\left( s - {s'} \right)  - {s'}\,\left(
         2\,s\,{\Sigma} - 3\,s\,{s'} - 2\,{\Sigma}\,{s'} + {{s'}}^2
       \right)  + 4\,{s'}\,\left( s' - s \right) \,{q_{s'}}^2
     \right)}{2\,\pi \,\left( s - {s'} \right)
     \,{{s'}}^3\,{q_{s'}}^2},\\[3mm]
   K^{(0)}_{1,0}(s,t') & = \frac{{t'}\,\left\{ \ln ({t'} + 4\,{q_{s}}^2)-
       \ln ({t'}) \right\}  - 2\,\left\{ 2 + \ln ({t'}) - \ln ({t'}
       + 4\,{q_{s}}^2) \right\} \,{q_{s}}^2}{8\,{\sqrt{3}}\,\pi
     \,{q_{s}}^4},\\
   G^{+}_{0,0}(t,s') & =  \frac{\sqrt{3}}{\pi}\,\left( \frac{\Sigma-s'}
       {2 s' q_{s'}^2} + 
      \frac{4}{{\sqrt{4\,m^2 - t}}\,
         {\sqrt{4\,M^2 - t}}}\,\,{\rm arccoth}\left(\frac{t-2\,
              \left( \Sigma - {s'} \right) 
              }{{\sqrt{4\,m^2 - t}}\,
             {\sqrt{4\,M^2 - t}}}\right) \right),\\[3mm]
   G^{+}_{0,1}(t,s') & =  \frac{3\,{\sqrt{3}}\,\left( \Sigma - {s'} - 
       t \right) }{2\,\pi \,s' q_{s'}^2 } + \frac{6\,{\sqrt{3}}\,
     \left( 2 s' q_{s'}^2 + \,{s'}\,t \right) }
     {\pi \, s' q_{s'}^2\,{\sqrt{4\,m^2 - t}}\,
     {\sqrt{4\,M^2 - t}}}  \,
     {\rm arccoth}\left(\frac{t-2\,
          \left( \Sigma - {s'} \right)}
         {{\sqrt{4\,m^2 - t}}\,{\sqrt{4\,M^2 - t}}}\right),\\[3mm]
   G^{(0)}_{0,0}(t,t') & = \frac{t}{\pi t'(t'-t)},\\[5mm]
   K^{-}_{0,0}(s,s') & = \frac{-\left( 3\,s^3 - 3\,s^2\,{s'} +
       2\,s\,{{s'}}^2 + s\,{\Delta }^2 + {s'}\,{\Delta }^2 -
       2\,s\,\left( s + {s'} \right) \,\Sigma        \right) }{2\,\pi
     \,s\,\left( s - {s'} \right) \,\left( {{s'}}^2 + {\Delta }^2 -
       2\,{s'}\,\Sigma  \right) }\\
   &  \quad + \frac{s\,\left\{ \ln (s\,{s'} -
       {\Delta }^2) - \ln (s\,\left[ s + {s'} - 2\,\Sigma \right] )
     \right\} }{\pi \,\left( s^2 + {\Delta }^2 - 2\,s\,\Sigma  \right)
     },\\[3mm]
   K^{-}_{0,1}(s,s') & = \frac{-3\,\left\{ {\Delta }^2 + s\,\left(
         3\,s + 2\,{s'} - 2\,\Sigma \right)  \right\} }{2\,\pi
     \,s\,\left( {{s'}}^2 + {\Delta }^2 - 2\,{s'}\,\Sigma  \right) }\\
   & \quad -\frac{3\,s\,\left( {s'}\,\left( 2\,s + {s'} -
         2\,\Sigma  \right)-{\Delta }^2 \right) \,\left\{ \ln (s\,{s'} - {\Delta
         }^2) - \ln (s\,\left[ s + {s'} - 2\,\Sigma \right] ) \right\}
     }{\pi \,\left( s^2 + {\Delta }^2 - 2\,s\,\Sigma  \right) \,\left(
       {{s'}}^2 + {\Delta }^2 - 2\,{s'}\,\Sigma  \right) },\\[3mm]
   K^{(1)}_{0,1}(s,t') & = \frac{3\,\left\{ {\Delta }^2 - s\,\left(
         3\,s + 2\,{t'} - 2\,\Sigma \right)  \right\}
     }{8\,{\sqrt{2}}\,\pi \,s\,{t'}}\\
   & \quad - \frac{3\,s\,\left( 2\,s + {t'}
       - 2\,\Sigma  \right) \,\left\{ \ln (-2\,s\,{t'}) - \ln
       (-2\,\left[ {\Delta }^2 + s\,\left( s + {t'} - 2\,\Sigma
         \right) \right] ) \right\} }{4\,{\sqrt{2}}\,\pi \,\left( s^2 +
       {\Delta }^2 - 2\,s\,\Sigma  \right) },\\[3mm]
   K^{-}_{1,0}(s,s') & = \frac{-\left\{ s^4 + {\Delta }^4 - 4\,s^3\,\Sigma  - 
       4\,s\,{\Delta }^2\,\Sigma  + 2\,s^2\,\left( 6\,{{s'}}^2 +
         7\,{\Delta }^2 - 12\,{s'}\,\Sigma  + 2\,{\Sigma }^2 \right)
     \right\} }{6\,\pi \,s\,\left( s^2 + {\Delta }^2 - 2\,s\,\Sigma
     \right) \,\left( {{s'}}^2 + {\Delta }^2 - 2\,{s'}\,\Sigma
     \right) }\\ 
    & \quad + \frac{s\,\left\{ \Delta^2 - s\,\left( s + 2\,{s'} -
          2\,\Sigma  \right) \right\} \,\left\{ \ln (s\,{s'} - {\Delta
          }^2) - \ln (s\,\left[ s + {s'} - 2\,\Sigma  \right] )\right\}
      }{\pi \,{\left( s^2 + {\Delta }^2 - 2\,s\,\Sigma  \right) }^2},\\[3mm]
   K^{-}_{1,1}(s,s') & = \frac{\left( s + {s'} \right) }
   {2\,\pi \,s' \left( s' - s \right)} + \frac{6\,
     \left( s^2 + s\,{s'} \right) }{\pi \,
     \left( s' - s \right) \,
     \left( s^2 + \Delta^2 - 2\,s\,\Sigma  \right) } + 
  \frac{s^3\,{s'} - 24\,s^2\,{{s'}}^2 - 
     s\,{{s'}}^3 - s^2\,{\Delta }^2 + 
     {{s'}}^2\,{\Delta }^2}{2\,\pi \,s\,s'\,
     \left( s' - s \right)\,
     \left( {s'}^2 +\Delta^2 - 
       2\,{s'}\,\Sigma  \right) }\\
   & \quad + 
  \frac{3\,s\,\left\{ \Delta^2 - 
       {s'}\,\left( 2\,s + {s'} - 
          2\,\Sigma  \right)  \right\} \,
     \left\{ \Delta^2 - 
       s\,\left( s + 2\,{s'} - 2\,\Sigma  \right) 
       \right\} \,\left\{ \ln (s\,{s'} - 
         {\Delta }^2) - \ln (s\,
         \left[ s + {s'} - 2\,\Sigma  \right] )
       \right\} }{\pi \,{\left( s^2 + {\Delta }^2 - 
         2\,s\,\Sigma  \right) }^2\,
     \left( {{s'}}^2 + {\Delta }^2 - 
       2\,{s'}\,\Sigma  \right) },\\[3mm]
   K^{(1)}_{1,1}(s,t') & = \frac{-\left( s^4 + {\Delta }^4 + 
       4\,s^3\,\left( 6\,{t'} - \Sigma  \right)  - 
       4\,s\,{\Delta }^2\,\Sigma  + 
       2\,s^2\,\left( 6\,{{t'}}^2 + {\Delta }^2 - 
          12\,{t'}\,\Sigma  + 2\,{\Sigma }^2
          \right)  \right) }{8\,{\sqrt{2}}\,\pi \,s\,
     {t'}\,\left( s^2 + {\Delta }^2 - 
       2\,s\,\Sigma  \right) }\\
   & \quad - \frac{3\,s\,\left\{ {\Delta }^2 + 
       s\,\left( s + 2\,{t'} - 2\,\Sigma  \right) 
       \right\} \,\left\{ 2\,s + {t'} - 
       2\,\Sigma  \right\} \,
     \left\{ \ln (-2\,s\,{t'}) - 
       \ln (-2\,\left[ {\Delta }^2 + 
           s\,\left( s + {t'} - 2\,\Sigma  \right) 
           \right] ) \right\} }{4\,{\sqrt{2}}\,\pi \,
     {\left( s^2 + {\Delta }^2 - 2\,s\,\Sigma  \right) }^2},\\[3mm]
   G^{-}_{1,0}(t,s') & = \frac{- {\sqrt{2}}\,\left\{
         8\,{{\Delta}}^2 + 4\,{{\Sigma}}^2 + 12\,{{s'}}^2 + t^2 -
         4\,{\Sigma}\,\left( 6\,{s'} + t \right)  \right\}
     }{3\,\pi \,\left\{ {{\Delta}}^2 + {s'}\,\left( -2\,{\Sigma} + {s'}
       \right)  \right\} \,{\sqrt{t+2\,{\Delta} - 2\,{\Sigma}
        }}\,{\sqrt{t-2\, {\Delta} -2\, {\Sigma}}}}\\
   &  \quad +\frac{4\,{\sqrt{2}}\,\left( 2\,{s'} + t -2\,{\Sigma} \right)
     }{\pi \,\left( 2\,\left( {\Delta} +
         {\Sigma} \right)  - t \right) \,{\sqrt{{\left( t+2\,{\Delta} -
             2\,{\Sigma} \right) }^2}}}\,
   {\rm arccoth}\left(i\,\frac{2\,{\Sigma} - 2\,{s'} -
       t}{{\sqrt{2\,{\Sigma}
           -2\,{\Delta} - t}}\,{\sqrt{-2\,{\Sigma} -2\, {\Delta}
          + t}}}\right),\\[3mm]
   G^{-}_{1,1}(t,s') & = - \frac{{\sqrt{2}}\,\left\{
         8\,{{\Delta}}^2 + 4\,{{\Sigma}}^2 + 12\,{{s'}}^2 +
         24\,{s'}\,t + t^2 - 4\,{\Sigma}\,\left( 6\,{s'} + t \right)
       \right\} }{\pi \,\left\{ {\Delta}^2 + {s'}\,\left( -2\,{\Sigma}
           + {s'} \right)  \right\} \,{\sqrt{t+2\,{\Delta} - 2\,{\Sigma}
          }}\,{\sqrt{t-2\,{\Delta}-2\, {\Sigma}}}}\\ 
&  \quad  + \frac{12\,{\sqrt{2}}\,\left( 2\,{s'} + t -2\,{\Sigma}
  \right) \,\left\{ {\Delta}^2 + {s'}\,\left( s' +
      2\,t -2\,{\Sigma}\right)  \right\} \,}{\pi \,\left\{
    {\Delta}^2 + {s'}\,\left(  s' -2\,{\Sigma} \right)  \right\}
  \,\left( 2\,\left( {\Delta} + {\Sigma} \right)  - t \right)
  \,{\sqrt{{\left( t+2\,{\Delta} - 2\,{\Sigma} \right) }^2}}}\times\\
& \qquad  {\rm arccoth}\left(i\,\frac{2\,{\Sigma} - 2\,{s'} -
    t}{{\sqrt{2\,{\Sigma} - 2\,{\Delta} - t}}\,{\sqrt{-2\,
          {\Sigma} -2 {\Delta}  + t}}}\right),\\[3mm]
   G^{(1)}_{1,1}(t,t') & = \frac{t\,{\sqrt{t-2\,{\Delta} - 2\,{\Sigma}
        }}\,{\sqrt{t+2\,{\Delta} - 2\,{\Sigma}}}}{4\,\pi
     \,{t'}\,\left( t' -t \right) }. 
\end{align*}
}
\end{appendix}


\begin{thebibliography}{99}

\bibitem{GLNPB}
J.~Gasser and H.~Leutwyler,
Nucl.\ Phys.\  {\bf B250} (1985) 465.


\bibitem{BKM}
V.~Bernard, N.~Kaiser and U.-G.~Mei{\ss}ner,
Nucl.\ Phys.\  {\bf B357} (1991) 129.

\bibitem{dobado} 
A.~Dobado and J.~R.~Pel{\'a}ez,
Phys.\ Rev.\  {\bf D56} (1997) 3057
[hep-ph/9604416].

\bibitem{Lang}
C.~B.~Lang,
Fortsch.\ Phys.\  {\bf 26} (1978) 509.


\bibitem{GLANN}
J.~Gasser and H.~Leutwyler,
Annals Phys.\  {\bf 158} (1984) 142.

\bibitem{KMSF1}
M.~Knecht, B.~Moussallam, J.~Stern and N.~H.~Fuchs,
Nucl.\ Phys.\  {\bf B457} (1995) 513
[hep-ph/9507319];
M.~Knecht, B.~Moussallam, J.~Stern and N.~H.~Fuchs,
Nucl.\ Phys.\  {\bf B471} (1996) 445
[hep-ph/9512404].

\bibitem{BCEGS}
J.~Bijnens, G.~Colangelo, G.~Ecker, J.~Gasser and M.~E.~Sainio,
Nucl.\ Phys.\  {\bf B508} (1997) 263
[hep-ph/9707291];
J.~Bijnens, G.~Colangelo, G.~Ecker, J.~Gasser and M.~E.~Sainio,
Phys.\ Lett.\  {\bf B374} (1996) 210
[hep-ph/9511397].

\bibitem{SSF} 
J.~Stern, H.~Sazdjian and N.~H.~Fuchs,
Phys.\ Rev.\  {\bf D47} (1993) 3814
[hep-ph/9301244].

\bibitem{AB1} 
B.~Ananthanarayan and P.~B\"uttiker,
Phys.\ Rev.\  {\bf D54} (1996) 1125
[hep-ph/9601285].


\bibitem{acgl}
B.~Ananthanarayan, G.~Colangelo, J.~Gasser, and H.~Leutwyler,
hep-ph/0005297.



\bibitem{clg}
G.~Colangelo, J.~Gasser, and H.~Leutwyler
Phys.\ Lett.\  {\bf B488} (2000) 261
[hep-ph/0007112].


\bibitem{Stern} J.~Stern,
hep-ph/9801282;
B.~Moussallam,
Eur.\ Phys.\ J.\  {\bf C14} (2000) 111
[hep-ph/9909292].

\bibitem{Dirac}
B.~Adeva et al., CERN/SPSC 2000-032, SPSC/P284 Add.~1

\bibitem{ulf_remark}
U.-G.~Mei{\ss}ner, private communication

\bibitem{Knecht:1993eq}
M.~Knecht, H.~Sazdjian, J.~Stern and N.~H.~Fuchs,
Phys.\ Lett.\  {\bf B313} (1993) 229
[hep-ph/9305332].

\bibitem{JN1}
N.~Johannesson and G.~Nilsson,
Nuovo Cim.\  {\bf A43} (1978) 376.

\bibitem{Steiner}
F.~Steiner,
Fortsch.\ Phys.\  {\bf 19} (1971) 115;
Fortsch.\ Phys.\  {\bf 18} (1970) 43


\bibitem{Roy}
S.~M.~Roy,
Phys.\ Lett.\  {\bf B36} (1971) 353.

\bibitem{Karabarbounis}
A.~Karabarbounis and G.~Shaw,
J.\ Phys.\ G {\bf G6} (1980) 583.


\bibitem{AB3}
B.~Ananthanarayan and P.~B\"uttiker, work in progress

\bibitem{Roessl}
A.~Roessl,
Nucl.\ Phys.\  {\bf B555} (1999) 507
[hep-ph/9904230].


\bibitem{NielsenOades}
H.~Nielsen and G.~C.~Oades,
Nucl.\ Phys.\  {\bf B55} (1973) 301

\bibitem{Hedegaard-Jensen}
N.~Hedegaard-Jensen,
Nucl.\ Phys.\  {\bf B77} (1974) 173

\bibitem{JohannessonandPetersen}
N.~O.~Johannesson and J.~L.~Petersen,
Nucl.\ Phys.\  {\bf B68} (1974) 397

\bibitem{Oller}
M.~Jamin, J.~A.~Oller and A.~Pich,
Nucl.\ Phys.\  {\bf B587} (2000) 331
[hep-ph/0006045].


\bibitem{Estabrooks:1978xe}
P.~Estabrooks, R.~K.~Carnegie, A.~D.~Martin, 
W.~M.~Dunwoodie, T.~A.~Lasinski and D.~W.~Leith,
Nucl.\ Phys.\  {\bf B133} (1978) 490.

\bibitem{Fuchs}
N.~H.~Fuchs, M.~Knecht and J.~Stern,
Phys.\ Rev.\  {\bf D62} (2000) 033003
[hep-ph/0001188].

\bibitem{BM}
B.~Ananthanarayan, P.~B\"uttiker, and U.-G.~Mei{\ss}ner, work in progress

\bibitem{BEG}
J.~Bijnens, G.~Ecker and J.~Gasser,
hep-ph/9411232, published in \cite{handbook}, pp. 125.

\bibitem{amoros}
G.~Amoros, J.~Bijnens and P.~Talavera,
Nucl.\ Phys.\  {\bf B585} (2000) 293
[hep-ph/0003258].




\bibitem{AB2}
B.~Ananthanarayan and P.~B\"uttiker,
Phys.\ Rev.\  {\bf D54} (1996) 5501
[hep-ph/9604217].


\bibitem{handbook}
L.~Maiani, G.~Pancheri and N.~Pancheri,
\newblock{\it The Second} DA$\Phi$NE {\it Physics Handbook}
(INFN-LNF-Divsione Ricerca, SIS-Ufficio Publicazioni, Frascati, 1995).

\end{thebibliography}
\end{document}